# Enhancing radical molecular beams by skimmer cooling

Hao Wu,*[a] David Reens,*[a] Tim Langen,†[a] Yuval Shagam,[a] Daniela Fontecha,[b] and Jun Ye[a]

A high-intensity supersonic beam source has been a key component in studies of molecular collisions, molecule-surface interaction, chemical reactions, and precision spectroscopy. However, the molecular density available for experiments in a downstream science chamber is limited by skimmer clogging, which constrains the separation between a valve and a skimmer to at least several hundred nozzle diameters. A recent experiment (*Science Advances*, 2017, **3**, e1602258) has introduced a new strategy to address this challenge: when a skimmer is cooled to a temperature below the freezing point of the carrier gas, skimmer clogging can be effectively suppressed. We go beyond this proof-of-principle work in several key ways. Firstly, we apply the skimmer cooling approach to discharge-produced radical and metastable beams entrained in a carrier gas. We also identify two different processes for skimmer clogging mitigation—shockwave suppression at temperatures around the carrier gas freezing point and diffusive clogging at even lower temperatures. With the carrier clogging removed, we now fully optimize the production of entrained species such as hydroxyl radicals, resulting in a gain of 30 in density over the best commercial devices. The gain arises from both clogging mitigation and favorable geometry with a much shorter valve-skimmer distance.

## 1 Introduction

Since a nozzle source was first proposed for producing monochromatic atomic and molecular beams in 1951,[1] supersonic beams have found widespread uses in a diverse set of research areas including atomic and molecular beam scattering,[2, 3] chemical reaction kinetics and molecular dynamics,[4-9] surface science,[10] helium droplets,[11] high-resolution spectroscopy,[12-14] etc. More recently, supersonic beams have been utilized together with a downstream molecular guide, for example a decelerator,[15-20] and the combination of these techniques has stimulated studies of molecular cooling and trapping,[21-23] cold collisions,[24, 25] precision measurement,[26, 27] and quantum effects in beam scattering and reactions.[28, 29]

There have been numerous improvements implemented in supersonic beam sources to boost initial densities.[30-33] However, the full performance[34] of a high density beam is limited by the formation of shockwaves near a conical collimating aperture known as a skimmer. Shockwaves are thin nonisentropic layers in a flow, with thicknesses on the order of several local mean free path lengths. The occurrence of shockwaves near a downstream surface is unavoidable in order for the leading edge of a continuum flow to match boundary conditions. Once the bulk of the flow encounters the shockwaves developed inside the skimmer, the beam transmission is greatly reduced. This phenomenon of dramatic beam transmission suppression is named "skimmer-clogging". In order to mitigate clogging, several key designs to improve the skimmer throughput have been implemented. The optimal angles for cone-shaped skimmers have been demonstrated for Campargue-type beam sources,[35, 36] which operate continuously with relatively high background pressure. For Fenn-type or pulsed sources in the rarefied regime, the clogging is less predictable as a function of cone angles, but still benefits from design optimization. The optimal parameters are determined by a compromise between a small external angle, which can prevent detached shockwaves, and a large internal angle, which minimizes the beam-wall collisions inside the skimmer. Further improvements are possible through more complex slit-skimmers.[37] However, even in these optimized designs, the valve-skimmer distance is still a critical parameter. In most experiments, the valve-skimmer distance has to be at least several hundred nozzle diameters[31] to avoid the formation of shockwaves. This large separation reduces, in an inverse squared manner, the density that can be loaded into a science chamber located after the skimmer. There is therefore an unavoidable trade-off between density reduction due to a large valve-skimmer distance and beam attenuation induced by clogging.

Recently, a new and very general technique—skimmer cooling—has been applied to pulsed beams and shown to significantly suppress skimmer-clogging for well-behaved carrier gases.[38] This can be explained intuitively as follows: once the surface of a skimmer is cold enough to adsorb the carrier gas particles without reflection upon contact, the surface boundary conditions for the flow are effectively removed to infinity, which guarantees that there will not be shockwaves. This cooling technique thus overcomes the density-limiting trade-offs.

Here, we demonstrate for the first time that the skimmer cooling technique works efficiently even for discharge-produced radical and metastable beams seeded in a carrier gas. A factor 30 gain of transmitted metastable neon (Ne*) is found by cooling the skimmer down to 8 K. Our results do not show any saturation and hence indicate that cooling to a lower temperature might further improve this gain. We believe the cooling technique is also suitable for all seeding experiments—whether discharge,[17] photolysis,[16] or ablation[13]— and thus allows chemically diverse molecular species to be generated with high density at the peak of a carrier gas pulse for the most efficient supersonic cooling. Moreover, we discover that two different clogging processes occur, depending on the temperature range. Finally, a factor 30 gain of hydroxyl radical (OH) density is demonstrated by a direct comparison between our 8 K skimmer and a well-optimized commercial room-temperature skimmer.

## 2 Experimental apparatus

Fig. 1 shows our experimental setup. The heart of this apparatus is a home-built, cryocooled skimmer, which has a 30° external angle and a 25° internal angle. The skimmer is indium-soldered onto a cold finger, which is thermally anchored to the 2nd stage of a 10 K pulse tube cryostat. A silicon temperature diode is installed several centimeters away from the base of the skimmer and a 20 W Nichrome-wire wrapped heater is bolted near the base of the skimmer to adjust the temperature. We have a two-step recipe for producing a low-temperature skimmer. First, both the skimmer and the cold finger are made of annealed 5N copper, which can provide a much higher thermal conductivity than OFHC.[39] Second, the majority of the cold finger is enclosed by a 70 K radiation shielding box made of OFHC copper, which minimizes the radiative heat load on the skimmer and the cold finger. Under this configuration, we are able to cool the skimmer to 8 K, which is confirmed with the temperature diode, even when the experiment is being run. The low temperature limit of 8 K is close to the no-load temperature of our cryostat, and we determine that the heat load to the skimmer is below 1 W and excellent thermal conduction is established. To study the temperature of the skimmer tip in detail, we performed a thermal modelling of the whole skimmer setup. Our results show a temperature difference of only 50 mK between the tip of the skimmer and the location of the temperature diode with a 1 W heat load.[40] Thus, the measured temperature should faithfully represent the real temperature of the skimmer.

Another benefit of skimmer cooling is that the skimmer acts as an efficient cryopump for the source chamber, reducing the background pressure by an order of magnitude to $10^{-7}$ Torr. This must be weighed against a potential drawback of skimmer cooling—the eventual limiting accumulation of ice. We observe no reduction in performance after a full hour of operation with 200 psi‡ stagnation pressure at 10 Hz repetition rate, but this may be different with a larger incident flux.

In our experiment, three different species are studied: neon, metastable neon (Ne*) and hydroxyl radicals (OH). A 30 µs long neon beam is produced by a room-temperature Even-Lavie valve in a non-clustering regime.[31] Ne* is generated by a dielectric barrier discharge (DBD) prior to a supersonic expansion. OH can be obtained through discharge of water vapor, which is provided by having water-soaked glass fiber filter papers installed inside the valve between the nozzle and a high-pressure neon gas cylinder. To stabilize the performance of the DBD, a tungsten filament is inserted into the source chamber to seed electrons towards the nozzle for discharge.

We use a variety of techniques to detect the three species under study. Neon traces after the skimmer are recorded by a fast ion gauge (FIG). A Mach-Zehnder (MZ) Interferometer[41] composed of a pair of backside polished mirrors is used to measure the neon density before the skimmer, not shown in Fig. 1. The MZ interferometer measurement shows, at the exit of the nozzle, neon has a peak density of $2\cdot10^{16}$ cm$^{-3}$ with 200 psi stagnation pressure. Ne* is detected by a microchannel plate detector (MCP). OH is probed with laser induced fluorescence (LIF). A 282 nm pulsed UV laser orthogonal to the molecular beam drives the transition from the ground state $(X\ ^2\Pi_{3/2}, J = 3/2, v = 0)$ to the excited electronic state $(A\ ^2\Sigma, N = 1, v = 1)$, and the resultant 313 nm fluorescence is focused by a pair of UV lenses onto a photomultiplier tube (PMT). In order to be sensitive only to the OH peak density but not the beam width, the detection volume is restricted to 1 mm$^3$ by the intersection of a 1.5 mm diameter laser beam and a 0.5 mm wide slit in a focal plane of the fluorescence collection system.

## 3 Experimental results and Data analysis

### 3.1 Neon

We begin with our results for the neon carrier gas, which confirms the efficacy of skimmer cooling as reported in ref. [38]. A factor of 9 peak signal gain is achieved during cooling from 35 K to 8 K. As shown in Fig. 2(a), at and above 35 K, only the leading edge of the gas pulse gets transmitted before the formation of shockwaves. In contrast, at 8 K a nearly Gaussian-shaped gas pulse is observed, which indicates clogging mitigation. The peak arrival time at 8 K is consistent with a speed of 790 m/s, the expected isenthalpic expansion speed of room temperature neon.

To further understand the extent of clogging mitigation, we investigate two different ways to vary the incident beam flux. When shockwaves are formed inside the skimmer, the clogging effect would worsen with a higher incident flux. One way to achieve a higher incident flux is to increase the stagnation pressure. As shown in Fig. 2(b), a ratio of neon before and after

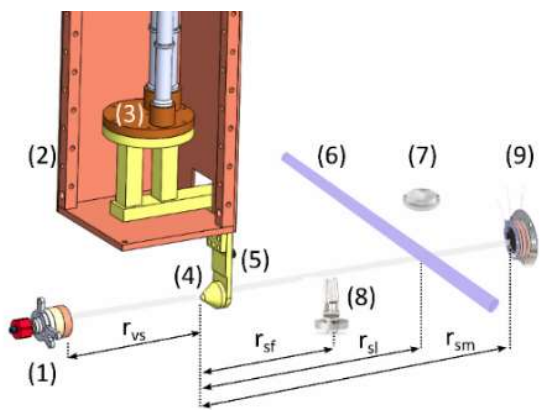

**Fig. 1** Schematic diagram, not to scale. (1) Even-Lavie valve. (2) 70 K radiation shield, of which two side panels are not shown. (3) 2nd stage of Cryomech PT807 10 K cryostat. (4) home-built conical copper skimmer. (5) Lakeshore DT-670 silicon temperature diode used for measuring the skimmer temperature. (6) 282 nm pulsed UV laser. (7) LIF collection lens. (8) Fast ion gauge (FIG). (9) Microchannel plates (MCP). $r_{vs}$ is the distance between the valve and the skimmer. $r_{sf}$ is the distance between the skimmer and the FIG. $r_{sl}$ is the distance between the skimmer and the laser. $r_{sm}$ is the distance between the skimmer and the MCP.

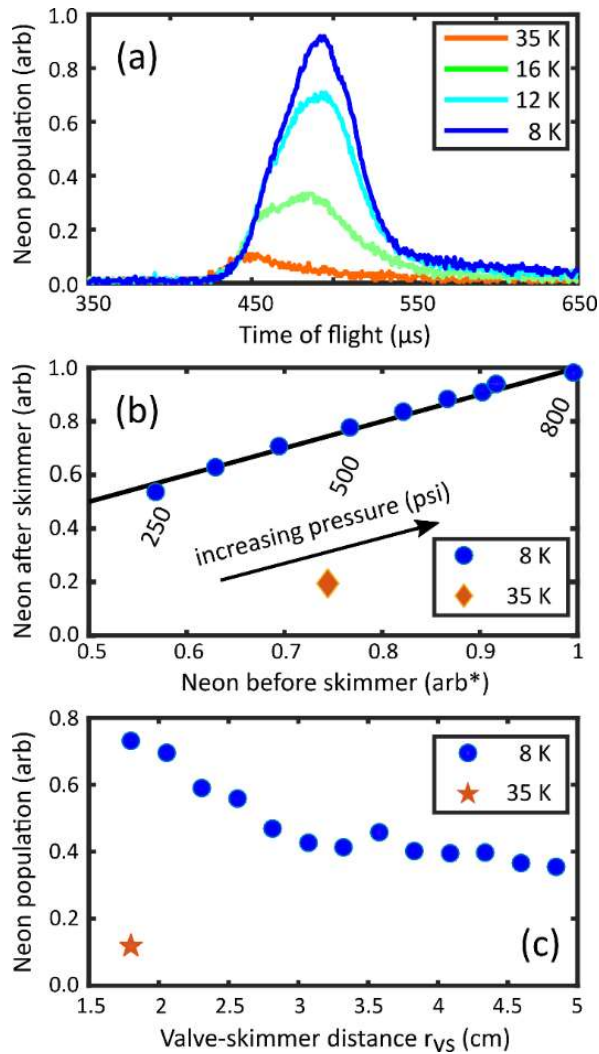

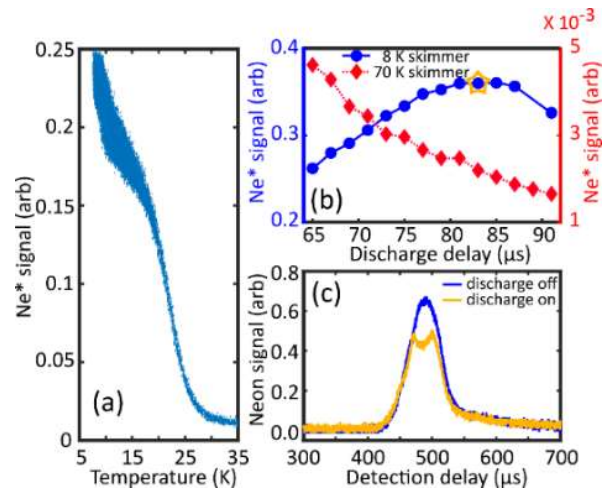

**Fig. 2** (a) Neon throughput for varying values of the conical skimmer temperature. The stagnation pressure is 400 psi‡ for panel (a) and $r_{vs}$ = 3 cm for panels (a-b). The transmitted Neon is measured at $r_{sf}$ =36 cm for panels (a-c). (b) Peak Neon signal before and after the skimmer at various stagnation pressures between 250-800 psi. The black solid line is a linear fit through the origin. A data point taken at 35 K (orange diamond) is included for comparison. (c) Peak Neon signal at different $r_{vs}$ with stagnation pressure 200 psi. A data point taken at 35 K (orange star) is shown for comparison.

the skimmer, which is independent of flux, suggests complete clogging mitigation at 8 K. The other way to vary the incident flux to the skimmer is by changing $r_{vs}$ (the distance between the valve and the skimmer). A continuing rise of the signal at smaller values of $r_{vs}$, even down to 2 cm, also confirms clogging mitigation (see Fig. 2(c)).[42]

### 3.2 Metastable neon

In a next step, we investigate the behavior of the cold skimmer using Ne*. For Ne*, we observe an even stronger signal increase by 30-fold during skimmer cooling to 8 K (see Fig. 3(a)). Moreover, the results indicate that a lower skimmer temperature could potentially lead to an even larger gain. The extra gain of Ne* relative to neon can be attributed to the

**Fig. 3** (a) Metastable Neon (Ne*) peak signal vs conical skimmer temperature. The stagnation pressure is 200 psi. $r_{vs}$ = 1.8 cm for panels (a-c). The discharge delay is fixed at 83 μs. Each shot of experiment is reflected as a point in the plot.[43] (b) Transmitted Ne* population vs. the discharge delay under two different temperatures. Ne* is seeded in the beam via dielectric barrier discharge (DBD) and detected at $r_{sm}$ = 160 cm. The DBD is composed of 17 cycles at 800 kHz. The stagnation pressure is 350 psi for panels (b-c). The delays here are measured relative to the valve firing for panels (b-c). (c) Neon pulses measured by FIG at $r_{sf}$ = 36 cm with the discharge on or off at 8 K. The discharge has an 83 μs delay relative to the valve firing, starred in panel (b). This optimum Ne* discharge timing occurs at the center of the neon beam, as evidenced by the clear depletion right at the peak position.

variation of optimal discharge timing as a function of temperature (and hence the degree of clogging). To achieve a maximal yield, the discharge timing should coincide with the peak of a carrier gas pulse. However, in the presence of clogging, only the front part of the carrier pulse would be able to go through the skimmer effectively before the skimmer is clogged (see the neon pulse comparison between 35 K and 8 K in Fig. 2(a)). Hence, the optimal discharge timing for the clogged beam must be set earlier than that in an unclogged one, to match the clogging-induced effective peak shift (see Fig. 3(b)). Only when the clogging is mitigated can we operate the discharge at its optimal timing coinciding with the peak of the carrier pulse.

This intuitive picture can be confirmed by examining the location of a discharge-induced depletion under the envelope of the neon carrier gas. We do this by taking FIG time of flight profiles of neon at $r_{sf}$ = 36 cm with the discharge toggled on or off. Fig. 3(c) shows this for the optimal discharge timing of 83 μs, starred in Fig. 3(b). It is seen that the Ne* is indeed produced right at the center of the neon packet.

Not only is the highest density achieved by seeding species at the peak of a carrier gas pulse, the most efficient supersonic cooling also occurs at the peak. We confirm this by fitting Gaussian distributions to the flight profiles of Ne* and extracting longitudinal temperatures. It is found that a Ne* beam as cold as 180 mK can be produced with the optimal 83 μs discharge delay. For comparison, the temperature increases by 40% to 260 mK with a smaller delay of 65 μs.

### 3.3 Shockwaves and diffusive clogging

We now explore clogging mitigation during skimmer cooling in more detail, and uncover a transition between two regimes. Our approach is empirical—we extract information about the nature of the clogging from the shape of the transmitted beam, where shape refers to its time of flight profile at the detector. As a figure of merit, we introduce the beam shape $\xi$—which compares the time of flight profile $w_T(t)$ at temperature $T$ to the Gaussian-shaped, unclogged profile $w_G(t)$ observed at 8 K:

$$\xi(T) = \frac{\int w_T(t) \cdot w_G(t) dt}{\sqrt{\int w_T^2(t) dt \cdot \int w_G^2(t) dt}} \quad (1)$$

When $\xi = 1$, $w_T$ and $w_G$ are identical up to a linear scaling; any difference in their shapes reduces the value of $\xi$ below unity. As shown in Fig. 2(a), for neon observed by FIG we find $\xi(35\ \text{K}) = 0.6$, corresponding to a vastly different profile, while $\xi(12\ \text{K})$ is nearly unity. We can also use $\xi$ to study the beam shape observed by MCP for Ne* at different skimmer temperatures, see Fig. 4(a). The time-of-flight profiles of Ne* require additional interpretation related to the double peak

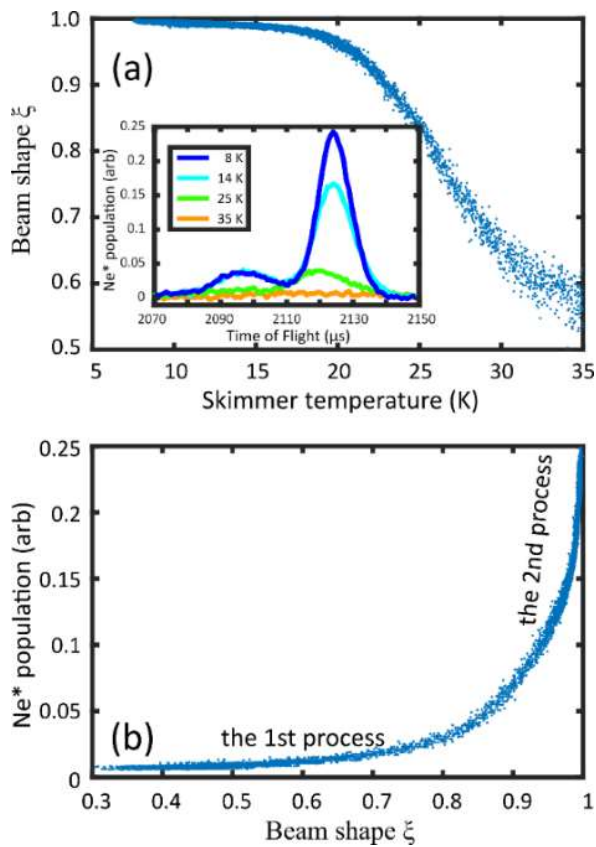

**Fig. 4** (a) The beam shape of Ne* vs conical skimmer temperature. Beam shape $\xi$ is defined as how close the time of flight profile at a certain temperature is to the unclogged and nearly Gaussian profile observed at 8 K. The inset panel shows transmitted Ne* beams at different skimmer temperatures. The double peak structure is related to minority species generated during the discharge, see the main text. (b) The Ne* peak signal vs $\xi$. From the bottom left to the top right, the temperature varies from 35 K to 8 K. The stagnation pressure is 200 psi and $r_{vs}$= 1.8 cm for panels (a-b). Each shot of experiment is reflected as a point in the plot.

structure shown in the inset. We associate the pre-peak with Rydberg neon species that are field ionized and accelerated into the detector ahead of the Ne*. We confirm this by increasing the voltage of the front plate of the MCP—which dramatically enhances the pre-peak and leaves the second unaffected. To calculate $\xi$ for Ne*, we first use double Gaussian functions to fit the beam profile, and then we exclude the first Gaussian profile attributed to the Rydberg species.

Having established $\xi$, we now evaluate it for Ne* across all measured profiles during skimmer cooling from 35—8 K. As shown in Fig. 4(a), $\xi$ increases dramatically from 35—20 K but then levels off near unity well before the gains of transmitted Ne* population cease, see Fig. 3(a). This can be understood further by plotting $\xi$ directly against Ne* population as in Fig. 4(b). The concave shape suggests the existence of two distinct clogging processes as the skimmer is cooled down—during the first process the beam shape increases but without significant signal gain, and during the second process the signal continues to gain after the beam shape has mostly stabilized.

We can interpret the two processes as follows: the first process is the suppression of dispersive shockwaves. These shockwaves are an inevitable phenomenon when a continuum supersonic flow interacts with boundaries such as the skimmer tip. They extend across the beam and cause significant heating and beam shape deviation. As noted and directly imaged in ref. [38], skimmer cooling reduces the influence of these shockwaves primarily by adsorbing molecules that would have otherwise participated in the formation of shockwaves. The adsorption relaxes the mass flow continuity constraints for shockwave formation and reduces their influence until they are completely suppressed. This is evidenced by the lack of heating or beam shape deviation measured by our near-unity $\xi$ parameter below about 20 K.

The additional two-fold signal gain below 20 K is associated with the rarefied equivalent of a shockwave—particles that reflect from the skimmer and interfere with the beam but are nonetheless too rarified to form shockwaves. We refer to this as diffusive clogging, and further interpret it as follows: When molecules that reflect off of the skimmer pass through the beam with few enough collisions, shockwaves no longer form. These reflected molecules, even when fully accommodated to the cryocooled but stationary skimmer, have hundreds of Kelvin worth of collision energy relative to the fast, supersonically cooled beam. Therefore, collisions between reflected molecules and beam molecules result in pairs that are still very hot relative to the beam. In the shockwave regime, these pairs collide further until all of their energy is dissipated into the beam, leading to the beam heating discussed above; but in the diffusive regime, they stop colliding while still hot. Thereafter, they rapidly diffuse relative to the cold centerline beam and are not detected. In this manner, the beam retains its cold temperature and near-unity $\xi$ parameter despite population loss. The transition between these regimes should correspond with the expected number of collisions approaching unity. Specifically, the mean free path $\lambda$ of beam molecules into reflected molecules (or their daughter pairs) is comparable to the length-scale L of the skimmer tip region relevant to shockwave formation. Throughput across this region should then follow Beer's law—with the fraction passing unperturbed given by $e^{-L/\lambda} \sim 1/e$. This leaves a factor of $e$ to be gained by further suppression of diffusive clogging.

Therefore, this simple model—shockwave suppression due to rarefaction when the mean-free path ratio reaches unity—explains both the observed beam shape behavior and the large gain remaining in the diffusive clogging regime. An additional corollary to this continued diffusive clogging is that without perfect adsorption, skimmer shape still plays a role, since a small external angle and a sharp tip reduce the ability of molecules to interfere in the diffusive clogging manner. In preliminary experiments with a thicker, 70° external angle skimmer, we found less optimal results than with the 30° skimmer used for all data reported here.

### 3.4 OH radical density comparison between two optimized skimmers

It is now clear that skimmer cooling can mitigate both shockwaves and diffusive clogging, but a key question is whether this method really represents an absolute improvement relative to the previous state of the art. To address this, we perform an OH density comparison between two optimized skimmers—a 300 K commercial skimmer and an 8 K skimmer (see Fig. 5). The LIF laser is located reasonably close to the skimmer ($r_{sl}$ = 6.6 cm). Our results show a factor of 30 gain achieved by skimmer cooling. In this region, we expect two types of gain. The first would be the geometric gain resulting from a reduced valve-detector distance. This gain can be further separated into transverse and longitudinal contributions. Assuming that the transverse density expansion follows $1/r^2$ position dependence in the free flight regime, the expected transverse contribution is a factor of $(20.9\ cm/8.4\ cm)^2$ = 6.2. The longitudinal expansion contributes to another factor of 8.9 µs/7.5 µs = 1.2, according to the FWHM of Gaussian fittings in Fig. 5. The second gain would be from actual clogging mitigation. This gain can be estimated by moving the laser and detection system to be far behind the skimmer ($r_{sl}$ = 70 cm), where the geometrical gain is negligible, and repeating the OH comparison between two skimmers. A population gain of 3.2 between 8 K and 300 K skimmers is found. Overall, the total expected gain is thus 6.2 x 1.2 x 3.2 = 24, which reasonably agrees with the measured factor of 30.

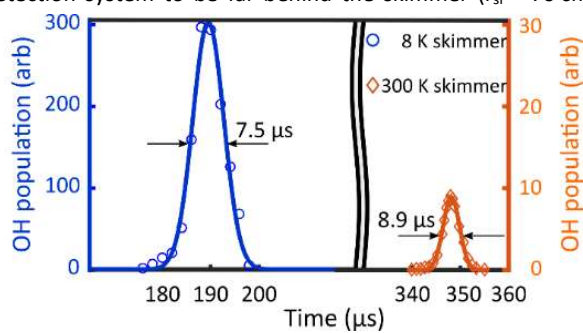

**Fig. 5** A direct comparison between a 300 K commercial skimmer and an 8 K home-made skimmer for the production of hydroxyl radical OH. The OH density is measured at a position behind skimmer ($r_{sl}$=6.6 cm) suitable for a molecular guide. The blue circle data is taken with an 8 K skimmer at a valve-laser distance of 8.4 cm and the orange diamond data is taken with a 300 K skimmer (Beam Dynamics model: 50.8) at a valve-laser distance of 20.9 cm. The solid lines are Gaussian fits for extracting the relative beam widths. The arbitrary scales for the left and right axes are in the same units. Time is recorded relative to valve firing.

To ensure that the commercial skimmer is actually well optimized, we see that the beam shape after the commercial skimmer is also near unity, confirming that there are no shockwaves developing. The skimmer position of $r_{vs}$=12 cm is experimentally selected for the optimum density and consistent with the recommended distance in ref. [31]. As has been discussed, we do expect to find an optimum that involves a trade-off between clogging and geometric density reduction.

## 4 Conclusions and outlook

We have demonstrated how skimmer cooling can lead to large gains for discharge-produced radicals and metastable species. Our results indicate that this technique can also be applied to many other species and production techniques. Moreover, our results reveal the existence of two distinct clogging mitigation processes. While the suppression of shockwaves dominates at moderately low temperatures, more efficient diffusive clogging mitigation can lead to further important gains in molecular density at even lower temperature. Notably, a factor of 30 gain in the OH density is achieved with an 8 K skimmer by combination of clogging mitigation and a smaller valve-skimmer distance. With this combination, a much brighter beam is available for a downstream molecular guide, such as our next generation Stark decelerator.[23] In such a setting our results bring a series of new questions, such as how to achieve an optimal mode matching between the brightened beam and the downstream molecular guide. However, there is no doubt that skimmer cooing will have an important impact on the large variety of experiments that rely on high molecular densities.

As far as other carrier gases are concerned, skimmer cooling could still be a general and feasible technique within a reasonable temperature range. It has been demonstrated that a skimmer temperature on the order of 10 K is sufficient for carrier gas heavier than neon due to their relatively high cryocondensation temperature.[38] Since lighter carrier gases, such as helium, can provide higher densities and more efficient cooling, it would be very important if this technique could also be extended to them. The challenge is that helium hardly condenses onto a copper surface above 1 K.[44] Nevertheless, skimmer cooling could still become feasible for helium in the 4 K regime with proper sorbents attached to the skimmer surface. It has been shown that with a µm-scale thickness pre-condensed Argon frost layer, the adsorption rate of helium/hydrogen can increase dramatically.[45] Also, simple porous sorbents such as activated charcoals[44] could lead to sufficient adsorption and hence unlock further unprecedented gains in density for future molecular beams.

## Acknowledgements

We acknowledge the Gordon and Betty Moore Foundation, the ARO-MURI, JILA PFC Phys-1734006, and NIST for their financial support. T.L. acknowledges support from the Alexander von Humboldt Foundation through a Feodor Lynen Fellowship. D.F. acknowledges support from the NSF REU Phys-1560023. We thank Y. Segev for helpful discussions. We also thank N. Punsuebsay for technical support.